\documentclass[12pt]{article}
\pdfoutput=1
\usepackage[utf8]{inputenc}
\usepackage[english]{babel}
\usepackage{amssymb,graphicx}
\usepackage{amsmath,amsfonts}
\usepackage{fullpage}

\def\ring#1{{\mathaccent'27 #1}}
\newcommand{\be}{\begin{equation}}
\newcommand{\ee}{\end{equation}}
\newcommand{\bea}{\begin{eqnarray}}
\newcommand{\eea}{\end{eqnarray}}

\newcommand{\g}{\gamma}

\newcommand{\bseq}{\begin{subequations}}
\newcommand{\eseq}{\end{subequations}}

\title{\vspace{-2cm} 
\begin{flushright}
{\normalsize
CERN-TH-2016-246} \\
\vspace{-0.5cm}
{\normalsize INR-TH-2016-044}
\end{flushright}
\vspace{0.5cm} 
Constraints on violation of Lorentz invariance from atmospheric
  showers initiated by multi-TeV photons}
\author{Grigory Rubtsov$^1$, Petr Satunin$^1$\thanks{{\bf e-mail}: satunin@ms2.inr.ac.ru}~, Sergey Sibiryakov$^{1,2,3}$
\vspace{.2cm}\\
\normalsize\it $^1$ Institute for Nuclear Research of the Russian Academy
of Sciences, \\  
\normalsize \it  60th October Anniversary Prospect, 7a, 117312  Moscow, Russia\\
\normalsize\it $^2$ Theoretical Physics Department, 
CERN, CH-1211 Gen\`eve 23, Switzerland\\
\normalsize\it $^3$ Institute of Physics, LPPC, \'Ecole Polytechnique 
F\'ed\'erale de Lausanne,\\[-0.05cm]
\normalsize\it CH-1015, Lausanne, Switzerland} 
\begin{document}
\date{}
\maketitle
\begin{abstract}
We discuss the effect of hypothetical violation of Lorentz invariance
at high energies on the formation of atmospheric showers by
very-high-energy gamma rays. In the scenario where Lorentz invariance
violation leads to a decrease of 
the photon velocity
with energy the formation of the showers is suppressed
compared to the Lorentz invariant case. Absence of such suppression in
the high-energy part of spectrum of the Crab nebula measured
independently by HEGRA and H.E.S.S. collaborations is used to set
lower bounds on the energy scale of Lorentz invariance
violation. These bounds are competitive with the strongest existing
constraints obtained from timing of variable astrophysical sources and
the absorption of TeV photons on the extragalactic background
light. They will be further improved by the next generation of multi-TeV
gamma-ray observatories.
\end{abstract}

\section{Introduction}

Very-high energy (VHE) gamma-ray astronomy is a rapidly developing
branch of astrophysics \cite{Rieger:2013rwa, Degrange:2015mda}. 
Since the first ground-breaking detection of 
several multi-TeV gamma-ray events from Crab nebula in
1989 \cite{Weeks}, it has evolved into
a well-established 
technique for high-quality astronomical observations. More than hundreds of TeV gamma-ray
sources have been discovered and studies of their spectra and
variability 
have made valuable contribution to our understanding 
of the internal processes in these objects
\cite{deNaurois:2015haa}. The propagation of VHE photons is affected
by the interstellar medium, in particular, the photon background and 
magnetic fields. Remarkably, it is also sensitive 
to tiny deviations from Lorentz invariance~(LI).

Possible violation of Lorentz invariance (or Lorentz violation (LV)
for short) is motivated by some approaches to quantum gravity 
(see reviews \cite{Mattingly:2005re,Liberati:2013xla} and references therein). 
Several approaches \cite{Ellis:2003if, 
  Girelli:2012ju, Horava:2009uw} predict that the
departures from LI, while being tiny at energies accessible in
laboratory, grow with energy and  
become significant at a certain high energy scale $M_{LV}$. This scale
is conventionally assumed to be of the order of Planck mass
$M_P=1.2\times 10^{19}$\,GeV, but can also lie a few
orders of magnitude
below\footnote{For example, non-projectable Ho\v rava-Lifshitz gravity
  \cite{Blas:2009qj} favors $M_{LV}$ in the range
  $10^{15} \div 10^{16}\, \mbox{GeV}$.}. 

LV 
in the non-gravity sector is conveniently parameterized within the
framework of effective field theory
\cite{Coleman:1997xq,Colladay:1998fq,Jacobson:2002hd,Myers:2003fd,
Kostelecky:2007fx}. 
In this framework one postulates existence of a
preferred frame, commonly identified with the rest-frame of the
CMB. Typical energy attained by a particle in astrophysical phenomena
is significantly higher (in CMB frame) than energy ever obtained in
the laboratory, making these phenomena a sensitive probe of 
LV \cite{Coleman:1997xq}. The most energetic 
particles detected in cosmic rays are hadrons (protons or nuclei)
with energies up to $10^{20}\,\mbox{eV}$
\cite{AbuZayyad:2012ru}. Their observation has
been used to set very stringent limits on LV for protons 
\cite{Gagnon:2004xh,Scully:2008jp,Bi:2008yx,Maccione:2009ju} and
nuclei \cite{Saveliev:2011vw}. However, hadrons are not elementary
particles and relating these bounds to the fundamental parameters of a
given model presents a complicated task. On the other hand, particles
from the sector of quantum electrodynamics (QED) --- photons,
electrons and positrons --- are elementary\footnote{This is true in the
  simplest setup assumed in this paper. In more complicated scenarios 
\cite{Kiritsis:2012ta,Bednik:2013nxa} the QED states, as well as
all other particles of the Standard Model, can be composite, which
suppresses observable effects of LV.} 
and constraints on their
properties translate directly into the constraints on the underlying
theory.   

The key consequence of LV is the change in particles' 
dispersion relations \cite{Myers:2003fd}. This has two potential
implications for VHE gamma rays. First, the dependence of the
photon propagation velocity on energy induces delays in the
arrival time of photons with different energies that can be
constrained by timing observations of variable distant
sources \cite{Biller:1998hg}. 
Second, it modifies the rates of particle reactions
\cite{Coleman:1997xq,Jacobson:2002hd,Colladay:2001wk,Rubtsov:2012kb}.  
Several
processes, such as photon decay $\gamma\to e^+e^-$ or photon splitting
$\gamma \to 3 \gamma$, kinematically forbidden in LI theory, can become
allowed. Cross-sections of other reactions, allowed also in the standard case
(pair production on soft photon background or in the Coulomb field of
a nucleus in the atmosphere), get modified. 
This affects the predictions for the gamma-ray spectra of 
astrophysical sources. The absence of deviations from the predictions
of the standard LI theory in the observed spectra 
establishes bounds on the parameters describing LV. 

VHE photons arriving to the Earth are detected through particle
showers that they produce in the atmosphere. The depth at which the shower
is initiated is determined by the cross section of the first
photon--nucleus interaction, the dominant channel being $e^+e^-$
production in the Coulomb field of the nucleus --- the Bethe--Heitler
process~\cite{Bethe:1934za}. As discussed in
\cite{Vankov:2002gt,Rubtsov:2012kb,Rubtsov:2013wwa}, the cross
section of the latter process sensitively depends on LV parameters in
the QED sector. In an interesting parameter range the shower formation
is suppressed compared to the LI case, leading to the
suppression of the detected photon flux. In this paper we emphasize
the role of this effect in setting the constraints on LV and derive
the bounds following from the absence of suppression in the measured
spectrum of the Crab nebula. 

The paper is organized as follows.
In Sec.~\ref{sec:2} we briefly describe the framework for parameterizing
deviations from LI in QED and review the existing constraints on the
LV parameters focusing on the case of quartic dispersion
relations. In Sec.~\ref{sec:3} we discuss the effect of LV on the
formation of an atmospheric shower by a VHE photon and derive the
corresponding constraints on the scale of LV in the photon dispersion
relation using the measurements of the Crab nebula spectrum by
HEGRA and H.E.S.S. collaborations. We also
estimate the reach of the Cherenkov Telescope Array (CTA) and future
extensive air shower arrays in improving these
bounds. Section~\ref{sec:4} is devoted to conclusions.


\section{Existing constraints on Lorentz violation}
\label{sec:2}

The generic effect of LV is the modification of particles' 
dispersion relations. Assuming spatial isotropy in the preferred
frame, particle energy $E$ depends only on the absolute value of 
momentum $p$ in that frame. 
At momenta smaller than the LV scale
$M_{LV}$ it can be expanded in powers of $p$. 
Focusing on the
QED sector and keeping up to quartic terms, 
one writes the dispersion relations for photons and
electrons/positrons: 
\begin{equation}
\label{disprels4}
E^2_\gamma=p_\g^2+\frac{\epsilon_\g p_\g^4}{M_{LV,\g}^{2}}, 
\qquad   E^2_{e}=m^2_e+p^2_e(1+\delta_e) +\frac{\epsilon_e p_e^4}{M_{LV,e}^{2}}\;,
\end{equation}
where $\epsilon_{\g,e}$ can take values $\pm 1$ and we allowed the
scales suppressing the quartic contributions for photons and
electrons/positrons to be different in general. Note that, 
without loss of generality, we have set the quadratic correction to
the photon dispersion relation to zero, so that the low-energy
velocity of photons is normalized to one; this can be always achieved
by an appropriate rescaling of the space- or time-coordinates\footnote{
We do not consider in this paper loop corrections to the dispersion
relations that can induce a 
logarithmic running of the coefficients in (\ref{disprels4}) with momentum.}.

We have not included cubic terms in (\ref{disprels4}).  
Within the effective field theory framework,
such terms would arise from $CPT$-odd contributions in the Lagrangian
\cite{Jacobson:2002hd,Myers:2003fd,Kostelecky:2008ts,Mattingly:2008pw}. 
Phenomenologically, they are strongly constrained with the
required suppression scale being well above the Planck
mass, see e.g. \cite{Maccione:2007yc}. In what follows we assume that
the underlying theory is $CPT$ invariant\footnote{While the $CPT$
  symmetry follows from LI, the 
  converse is not true: a theory can be $CPT$ invariant and Lorentz
  violating at the same time.}, so that cubic corrections to the
dispersion relations are absent. 

Finally, the expressions (\ref{disprels4}) implicitly assume that the
dispersion relations are the same for states with different
helicities. For photons, this is guaranteed by the $CPT$ symmetry. On
the other hand, for the fermionic states $CPT$ invariance only ensures
that the dispersion relation of electron with positive (negative)
helicity coincides with the dispersion relation of positron with negative
(positive) helicity. We take the equality of the dispersion relations of
electrons with opposite helicities as an additional simplifying
assumption. In principle, it can be ensured by
requiring that the QED sector is invariant under parity
\cite{Rubtsov:2012kb}, as it happens 
in the LI case. Our results will not depend on this
assumption. 

Note that the parameters in the dispersion relations (\ref{disprels4})
can be connected with the coefficients in the 
Lagrangian of LV QED in the parameterization of
\cite{Kostelecky:2009zp,Kostelecky:2013rta},
\be
\delta_e = -2\ring{c}^{(4)}_2, \qquad  
\frac{\epsilon_e}{M_{LV,e}^{2}} = -2\ring{c}^{(6)}_4,\qquad 
\frac{\epsilon_\g}{M_{LV,\g}^{2}} = -\frac{c^{(6)}_{(I)00}}{\sqrt{\pi}}.
\ee
We now review the constraints on these parameters.

\paragraph{A. Constraints on LV in electrons.}

The parameter $\delta_e$ affects the physics at low energies and can
be constrained using terrestrial experiments. The analysis of
radiation losses by the electron and positron beams at LEP gives 
\cite{Altschul:2010na},
\be
\label{deltaLEP}
|\delta_e|< 2\times 10^{-15}\;.
\ee 
The constraints on $M_{LV,e}$ come from the observation of the Crab
nebula spectrum in the energy range up to $0.1$\,GeV. The spectrum has
two peaks well described by the synchrotron-self-Compton model 
(see \cite{Harding:2015rta} for review). This requires presence of
electrons with energies up to $E_{{\rm max},e}\sim 10^3$\,TeV in the plasma
inside the nebula. They produce synchrotron radiation that corresponds
to the low-energy hump of the spectrum and rescatter it by the inverse
Compton process giving rise to the high-energy peak. Possible LV in
electrons would modify the intensity of the synchrotron radiation and
hence change the Crab 
spectrum~\cite{Altschul:2006pv,Liberati:2012jf}. This leads to the
following bound~\cite{Liberati:2012jf},
\be
\label{electrons}
M_{LV,e} > 2\times 10^{16} \, \mbox{GeV}.
\ee
This analysis is insensitive to LV in photons as the energy of the
synchrotron radiation (up to $0.1$\,GeV) is much smaller than the
energy of electrons.

Ref.~\cite{Liberati:2012jf} performs the analysis under the assumption
$\delta_e=0$. However, relaxing this assumption is not expected to
significantly change the constraint (\ref{electrons}). Further, it is
instructive to estimate the bound on $\delta_e$ that can be obtained
if the analysis is performed allowing for its non-zero values. The
relevant quantity for the synchrotron radiation is the deviation of
the group velocity of electrons from unity. Comparing the
contributions to the group velocity from the quadratic and quartic
terms in the electron dispersion relation, we find that the bound
(\ref{electrons}) can be translated into (cf.~\cite{Stecker:2013jfa}),
\be
\label{deltaCN}
|\delta_e|\lesssim 3 (E_{{\rm max},e}/M_{LV,e})^2\sim 10^{-20}\;.
\ee 
Of course, this is only a crude estimate and a careful analysis taking
into account the dynamical processes in the Crab nebula is required to
set rigorous bounds on $\delta_e$. The fact that (\ref{deltaCN}) is
more than five orders of magnitude stronger than the best laboratory
constraint (\ref{deltaLEP}) makes such analysis promising. However, it
is beyond the scope of the present paper.

We are going to see that the constraints on LV in the photon
dispersion relation that can be obtained from the current data are
significantly weaker than for electrons. Therefore we will neglect LV
in electrons from now on.

\paragraph {B. Photon time of flight from distant sources.}
A quartic correction in the dependence of photon energy on momentum implies the
dependence of photon phase and group velocities on energy. Depending
on the sign of the correction, high-energy photons from
fast flares in distant sources would arrive earlier or later than
low-energy 
ones. The time of flight analysis has been performed for 
active galactic nuclei (AGN)~\cite{HESS:2011aa}, 
gamma-ray bursts (GRB)~\cite{Vasileiou:2013vra} and
pulsars~\cite{Zitzer:2013gka}. Absence of statistically significant
time-lags between photons with different energies yields,  
\begin{align}
\label{tof-1}
&M_{LV,\g} > 6.4\times 10^{10}\,\mbox{GeV},
&\mbox{AGN~\cite{HESS:2011aa}}\;,\\ 
\label{tof-2} 
&M_{LV,\g} > 1.3\times 10^{11}\,\mbox{GeV} 
&\mbox{GRB~\cite{Vasileiou:2013vra}}\;.
\end{align}
The bound from pulsars is significantly weaker. 

The time of flight bounds have the advantage of directly constraining
the photon dispersion relation, independently of the effects of LV on
the interactions. However, they are somewhat sensitive to the model of
the source flare that contributes the largest uncertainty in the
analysis. Stronger bounds on $M_{LV,\g}$ are obtained by considering
the physical processes affecting the propagation and detection of
VHE photons. The relevant processes differ depending   
on whether $\epsilon_\g$ is positive or
negative. With some abuse of language, we will refer to these cases as
``superluminal'' and ``subluminal'' respectively.

\paragraph {C. Photon decay to $e^+e^-$ pair.}
In the superluminal case ($\epsilon_\g=+1$) a high-energy photons can
decay into $e^+e^-$ pairs in the vacuum. This process occurs only if
the photon energy exceeds a certain threshold that can be found as
follows. The quartic contribution to the dispersion relation can be
thought of as an effective momentum-dependent ``photon mass'',
\be
\label{mgamma}
m^2_{\g,eff}(p_\g)\equiv E_\g^2-p_\g^2=\frac{p_\g^4}{M_{LV,\g}^2}\;.
\ee 
It characterizes the amount of energy that can be transferred from the
photon to the decay products. The process $\g\to e^+e^-$ becomes
allowed once $m_{\g,eff}$ exceeds\footnote{Recall that we neglect LV
  in electrons.} $2m_e$. The pair is created with approximately 
equal momenta ---
half of the initial photon momentum. Above the threshold the decay is
very rapid\footnote{When $m_{\g,eff}\gg 2m_e$ the decay width is given
by
$\Gamma_{\g\to e^+e^-}=(\alpha\, p_\g^3)/(3 M_{LV,\g}^2)$,
where $\alpha$ is the fine structure constant~\cite{Rubtsov:2012kb}.} 
and leads to a sharp cutoff in photon spectrum of all
astrophysical sources: no high-energy photons can reach the Earth from
astronomical distances~\cite{Jacobson:2002hd}. Thus, an observation of
gamma rays of astrophysical origin with an energy $E_\g$ gives the
bound,
\be
\label{photondecaybound}
M_{LV,\g} > \frac{E_\gamma^2}{2m_e}\;.
\ee
The recent analysis~\cite{Martinez-Huerta:2016azo} 
using the highest-energy photons observed from
the Crab nebula sets the constraint,
\be
\label{decayrecent}
M_{LV,\g} > 2.8\times 10^{12}\,\mbox{GeV}\qquad\qquad\qquad\qquad
(\epsilon_\g=+1).
\ee

Even if the photon decay into $e^+e^-$ is kinematically forbidden, the
flux from astrophysical sources can be depleted by photon splitting
$\g\to n\,\g$. This process is kinematically allowed whenever the
photon dispersion relation is superluminal. Splitting into 3
photons\footnote{The width of splitting into 2 photons $\g\to\g\g$
  is generally expected to be more suppressed by additional powers of
  the LV scale as in the limit of LI QED the matrix element with an
  odd number of external photon legs identically vanishes (this is the
  statement of the Furry theorem), see a discussion
  in~\cite{Liberati:2013xla}.}
$\g\to 3\g$ was analyzed in~\cite{Gelmini:2005gy} for the case of
cubic corrections to the photon dispersion relation and the width of
this process was found to strongly depend on energy and the LV
scale. Thus, observations of multi-TeV photons of astrophysical origin
put restrictive bounds on the latter. However, a study for the case of
quartic dispersion relation is missing in the literature. We leave the
derivation of the corresponding bounds for future. 

\paragraph {D. Modification of pair production on background photons.} 
Standard LI physics predicts that a VHE
photon interacts with extragalactic background light (EBL)
producing an $e^+e^-$ pair, $\gamma \gamma_b \to e^+e^-$, where $\gamma$
is the VHE photon and $\gamma_b$ is a photon from the background. The
mean free path of a photon with energy of several $\sim 100$\,TeV is 
$\sim \, 1\,\mbox{Mpc}$ \cite{Berezinsky:2016feh}, which leads to an
attenuation of VHE photon flux from extragalactic sources. For sources
within the Milky Way this process is irrelevant. 

Subluminal LV in photons ($\epsilon_\g=-1$) shifts the threshold of
pair production upward
\cite{Kifune:1999ex,Stecker:2001vb,Stecker:2003pw,Jacob:2008gj,Tavecchio:2015rfa}. 
This leads to higher predictions for the VHE photon flux from
extragalactic sources than in the LI case. Non-detection of large
fluxes constrains LV. 
Ref.~\cite{Biteau:2015xpa} uses the data on the Mrk 501 
flare in 1997 \cite{Aharonian:1999vy} to establish a bound on the
cubic correction to the photon dispersion relation. Translating it 
into the bound on the quartic term one obtains,
\be
\label{pp0bound}
M_{LV,\g} \gtrsim 3\times 10^{11}\,\mbox{GeV}\qquad\qquad\qquad\qquad
(\epsilon_\g=-1).
\ee
Recent analysis of the VHE part of the spectrum of Mrk 501 during the
2014 flare  
leads to a stronger limit \cite{Lorentz:2016aiz}:
\be
\label{ppbound}
M_{LV,\g} > 7.5\times 10^{11}\,\mbox{GeV}\qquad\qquad\qquad\qquad
(\epsilon_\g=-1)
\ee
at 95\% confidence level (CL). It is worth noting that these bounds rely on
the assumption that 
the observed cutoff in the Mrk 501 spectrum is not
intrinsic to the source, but is fully accounted for by absorption on
EBL. Besides, they require modeling of the EBL spectrum. While the
understanding of EBL has significantly improved over the last decade
(see \cite{Biteau:2015xpa,Stecker:2016fsg} and references therein),
some uncertainties still remain~\cite{Horns:2012fx,Rubtsov:2014uga}. 

In Refs.~\cite{Galaverni:2007tq,Maccione:2008iw} it was suggested that
a very strong constraint,
\be
\label{boundGZK}
M_{LV,\g}\gtrsim 1.2\times 10^{22}\,\mbox{GeV}\qquad\qquad\qquad\qquad
(\epsilon_\g=-1),
\ee
can be obtained from non-observation of a photon component in
ultra-high-energy (UHE) cosmic rays (energies $\gtrsim
10^{19}$\,eV). In the LI case photons with such energies get absorbed
through pair production on the cosmic microwave background (CMB),
whereas LV at a scale below (\ref{boundGZK}) would suppress this
process and UHE photons would reach the Earth.  
Clearly, this argument requires UHE photons to be
produced in the Universe in the first place. As such, it essentially
relies on the 
assumption that the dominant component of UHE cosmic
rays are protons that give rise to UHE photons through a cascade
starting with a pion production on CMB --- the GZK
process~\cite{Greisen:1966jv,Zatsepin:1966jv}. At the moment it is not
clear whether this assumption actually holds \cite{Abbasi:2015xga}.

\section{Effect of Lorentz violation on atmospheric showers}
\label{sec:3}

The bounds on $M_{LV,\g}$ reviewed in the previous section have been
derived under the assumption of standard interaction of high-energy
photons with the Earth' atmosphere. We now discuss the validity of
this assumption and obtain new constraints by considering the effect
of LV on photon-induced atmospheric showers.
We follow the approach of \cite{Rubtsov:2013wwa} which we adapt here
to the case of multi-TeV
energies.

\subsection {Suppression of the Bethe-Heitler process} 

A primary photon interacts with the atmosphere mainly through the
Bethe -- Heitler process --- pair production in the Coulomb field of
an atomic nucleus in the air. The standard LI result for the cross
section of this process reads \cite{Bethe:1934za}, 
\be
\label{BH}
\sigma_{\mathrm{BH}}=\frac{28Z^2\alpha^3}{9m_e^2}
\Big(\log \frac{183}{Z^{1/3}}-\frac{1}{42}\Big)\;,
\ee  
where $\alpha$ is the fine structure constant and $Z$ is the charge of
the nucleus; for scattering on nitrogen ($Z=7$) this gives
$\sigma_{\rm BH}\approx 0.51\,\mathrm{b}$.  
The depth of the first interaction $X_0$ is a random variable obeying
exponential distribution with the mean value 
$\langle X_0\rangle=m_{\rm at}/\sigma_{\rm BH}\,\simeq
57\,\mbox{g}\,\mbox{cm}^{-2}$, where $m_{\rm at}$ is the average mass of the
atoms of the air (typically, nitrogen). The first interaction leads to
the development of an electromagnetic cascade with the number of
particles in the cascade reaching its maximum at the depth 
$X_{\rm max}$. The length of the shower development 
$\Delta X\equiv X_{\rm max}-X_0$ follows the Gaussian statistics. The
mean value  
$\langle \Delta X\rangle$ depends logarithmically on the primary
photon energy and varies between $200\,\mbox{g}\,\mbox{cm}^{-2}$ and 
$250\,\mbox{g}\,\mbox{cm}^{-2}$ in the relevant energy range 
(from $100$\,GeV to $100$\,TeV). Within this range the dispersion
$\Sigma_{\Delta X}\approx 50\,\mbox{g}\,\mbox{cm}^{-2}$ is
approximately constant~\cite{Ohm}. 

As pointed out in \cite{Vankov:2002gt,Rubtsov:2012kb}, LV changes the
cross section of the Bethe -- Heitler process. Qualitatively this can
be understood as follows. The electron mass in the expression (\ref{BH})
characterizes the momentum transfer between the photon and nucleus
required to produce the $e^+e^-$ pair. In the LV case the momentum
transfer is shifted due to the presence of the effective photon mass
(\ref{mgamma}). Thus, up to a factor of order one, the modified Bethe
-- Heitler cross section can be estimated as (\ref{BH}) with the
replacement 
\be
\label{merepl}
m_e^2\mapsto |m_e^2-m_{\g,eff}^2(p_\g)/4|\;.
\ee 
This modification is not relevant for superluminal photons as the
cross section essentially remains close to its value in the LI theory
as long as $0<m^2_{\g,eff}(p_\g)<4m_e^2$, i.e. as long as the photon decay
is forbidden; for higher values of $m^2_{\g,eff}$ photon decay provides
the dominant signature of LV. However, for subluminal photons the
modification of the Bethe -- Heitler cross section can be
important. If
\be
\label{mesmall}
m^2_{\g,eff}(p_\g)<0~,~~~~~|m^2_{\g,eff}(p_\g)|\gg 4m_e^2
\ee
the cross section gets strongly suppressed.
These qualitative arguments are supported by an explicit calculation
in LV QED. Under the conditions (\ref{mesmall}) the modified cross
section reads~\cite{Rubtsov:2012kb},
\be
\label{BHLV}
\sigma_\mathrm{BH}^\mathrm{LV}=
\frac{16Z^2\alpha^3}{3|m^2_{\g,eff}(p_\g)|}\;\log\frac{1}{\alpha
  Z^{1/3}}\;\log\frac{|m^2_{\g,eff}(p_\g)|}{2m_e^2}\;. 
\ee
The suppression factor
\be
\label{suppr}
\frac{\sigma^{\rm LV}_{\rm BH}}{\sigma_{\rm BH}}\simeq 
\frac{12 m_e^2 M_{LV,\g}^2}{7E_\g^4}\cdot 
\log\frac{E_\g^4}{2m_e^2 M_{LV,\g}^2}
\ee
quickly decreases with energy.

Smaller cross section delays the formation of the electromagnetic
cascade which is now initiated deeper in the
atmosphere. Correspondingly, the depth of the maximal shower
development also increases. If it exceeds certain limiting value
$X_{\rm max}^{\rm lim}$ which depends on the experimental setup, the
event cannot be recognized as a photon. This implies a fast drop
in the number of registered photons above certain energy which is
determined by the LV scale $M_{LV,\g}$. Note that
this effect is opposite to the other consequence of LV discussed in
the previous section, namely, inefficient absorption of multi-TeV
photons on EBL which leads to the increase of the photon flux from
extragalactic sources. Therefore, in analyzing the constraints on LV 
it is important to make sure that
these two effects do not compensate each other.

\subsection{Constraints from observations of the Crab nebula}

\begin{figure}[t]
\hspace{-9mm}
\includegraphics[width=0.6\linewidth]{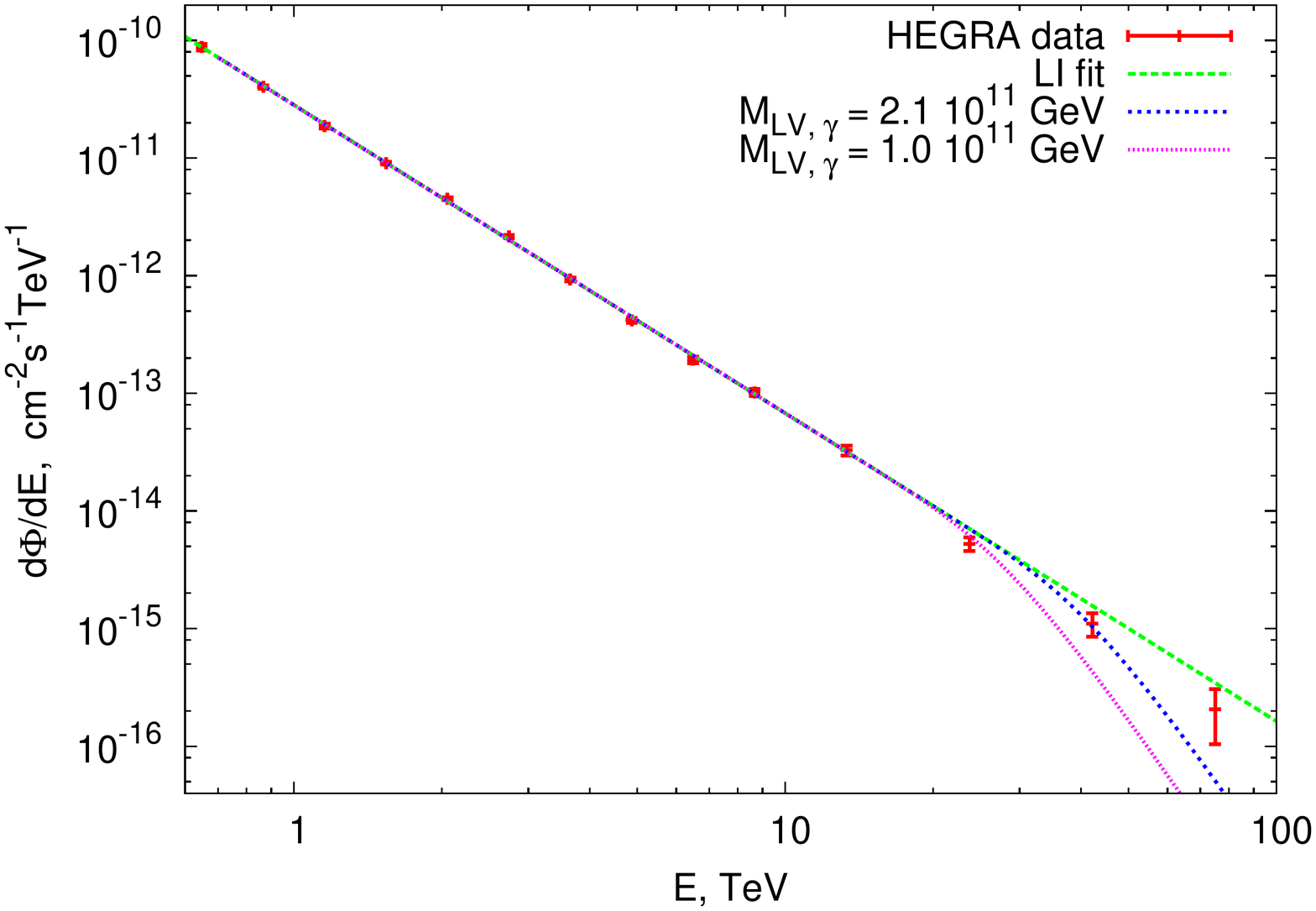}
\hspace{-13mm}
\includegraphics[width=0.6\linewidth]{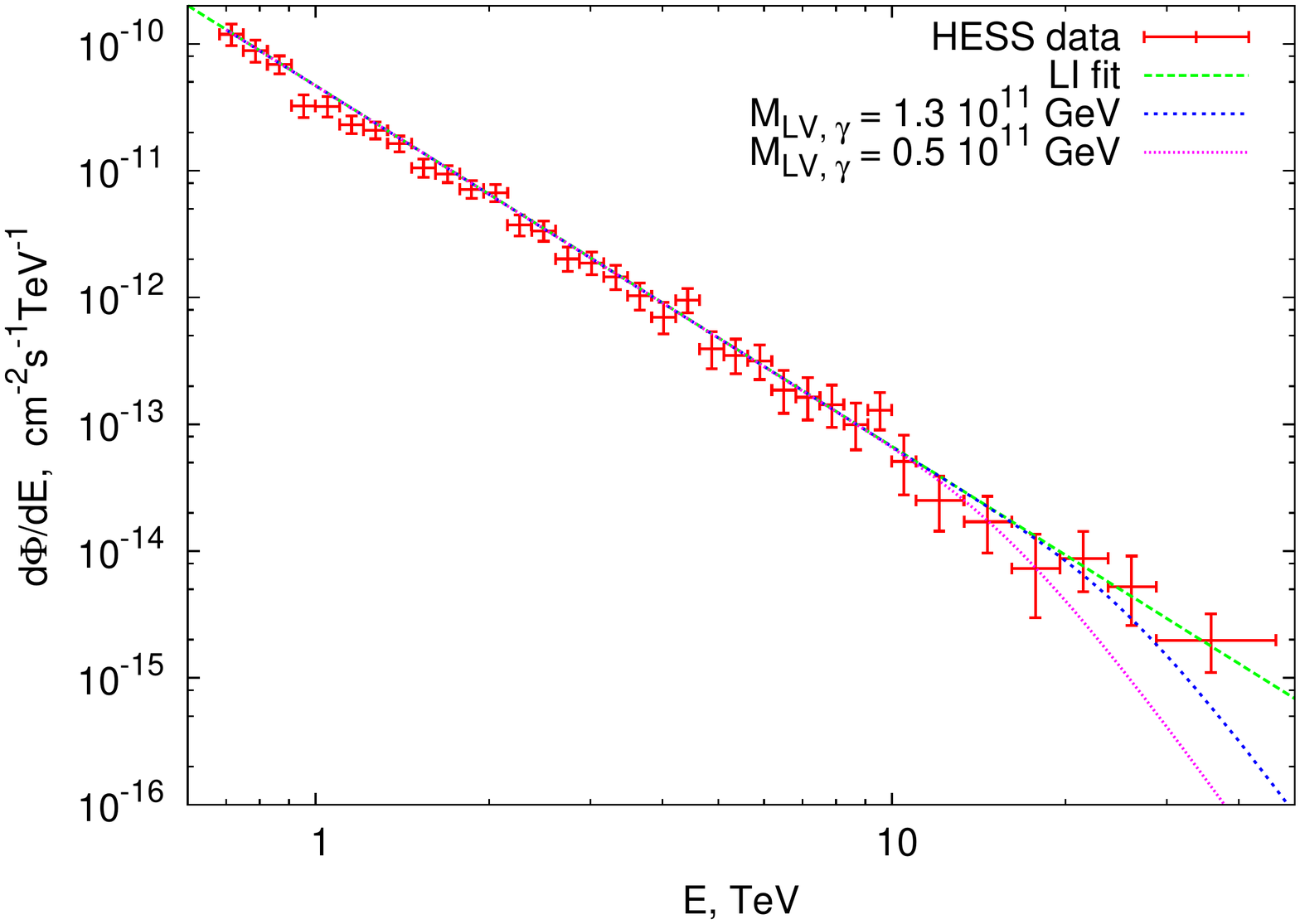}
\caption{\label{Fig1}
Photon spectrum of the Crab nebula obtained by collaborations
HEGRA (385 hours of data from 1997 to 2002)~\cite{Aharonian:2004gb} (left) and 
H.E.S.S. 
(4.4 hours of data during the flare of March 2013)~\cite{Abramowski:2013qea} 
(right). 
The dashed line corresponds to the best
power-law fit of the spectrum while the dotted lines show the
prediction for the flux under the hypothesis of Lorentz violation 
with a given $M_{LV,\g}$. 
}
\end{figure}

Absence of evidence for the suppression of the shower formation in the
observational data can be used to derive constraints on
$M_{LV,\g}$. The most energetic photon events have been detected from
the Crab nebula. The spectra were measured independently by the HEGRA
experiment up to $E_\g\sim 75$\,TeV \cite{Aharonian:2004gb} and by
H.E.S.S. up to  
$E_\g\sim 40$\,TeV \cite{Abramowski:2013qea}; they are shown in
Fig.~\ref{Fig1}. 
Both are well described
by a power law
\be
\label{plspectrum}
\bigg(\frac{d\Phi}{d E}\bigg)_{\rm obs}\propto E^{-n}~,
\qquad~~~
n=\begin{cases}
2.62\pm 0.02& \text{HEGRA \cite{Aharonian:2004gb}}\\
2.7\pm 0.1&\text{H.E.S.S. \cite{Abramowski:2013qea}}
\end{cases}
\ee
without any significant evidence for a cutoff. As the Crab nebula is a
galactic source, there is no significant absorption on EBL.

In the presence of LV the measured flux gets reduced,
\be
\label{LVhyp}
\bigg(\frac{d\Phi}{dE}\bigg)_{LV}= P_{\rm reg}(E_\g) 
\cdot \bigg(\frac{d\Phi}{dE}\bigg)_{LI}\;,
\ee
where $P_{\rm reg}(E_\g)$ is the probability to actually register a photon
with energy $E_\g$. The latter is equal to the probability that
$X_{\rm max}$ of the shower induced by the photon does not exceed
$X_{\rm max}^{\rm lim}$. To find $P_{\rm reg}$ we assume that
LV affects only the cross section of the first interaction and
does not modify the subsequent development of the shower. This is
justified as the secondary particles in the electromagnetic cascade
are less energetic than the primary photon. Then $P_{\rm reg}$ is
given by 
\be
\label{Preg}
P_{\rm reg}(E_\g)=\int_0^{X_{\rm max}^{\rm lim}}dX_{\rm
  max}\int_0^{X_{\rm max}}dX_0
\;\frac{{\rm e}^{-\frac{(X_{\rm max}-X_0-\langle\Delta
      X\rangle)^2}{2\Sigma_{\Delta X}^2}}}
{\sqrt{2\pi}\,\Sigma_{\Delta X}}\cdot
\frac{{\rm e}^{-X_0/\langle X_0\rangle_{LV}}}{\langle X_0\rangle_{LV}}\;,
\ee  
where
\be
\label{X0LV}
\langle X_0\rangle_{LV}=\frac{\sigma_{\rm BH}}{\sigma^{\rm LV}_{\rm BH}(E_\g)}
\langle X_0\rangle_{LI}\;.
\ee
The registration probability starts deviating significantly from unity
at the energies where $\langle X_0\rangle_{LV}$ becomes comparable
to $X_{\rm max}^{\rm lim}$. In the limit $\langle X_0\rangle_{LV}\gg
X_{\rm max}^{\rm lim}$ it tends to 
\be
P_{\rm reg}(E_\g) \simeq \frac{X_{\rm max}^{\rm lim}-\langle\Delta X\rangle}
{\langle X_0 \rangle_{LV}}\,
\ee
which reflects the fact that for large $\langle X_0 \rangle_{LV}$ the
probability to form a shower is uniformly distributed over the depth
of the atmosphere.

The effect of LV on the prediction for the Crab spectrum is
illustrated in Fig.~\ref{Fig1}. We take the primary spectrum to be
power-law with the spectral index fixed by the data points at 
energies below $20$\,TeV. One clearly sees a break in the
highest-energy tail of the spectrum predicted by the model due to
the suppression of shower formation. We now analyze the HEGRA and
H.E.S.S. datasets separately and obtain the constraints on $M_{LV,\g}$
from the excess of the number of actually observed events over that
predicted by the LV model.  

\paragraph{A. HEGRA data.}
HEGRA experiment has collected 385 hours of data of the Crab photon
spectrum in the multi-TeV range during the period from 1997 to
2002~\cite{Aharonian:2004gb}. The obtained spectrum shows power-law
dependence till the last energy bin\footnote{A slight
  steepening may be seen at the end of the spectrum, but its
  significance is less than $2\,\sigma$.} centered at $E_{\rm max}=75$\,TeV (see
Fig.~\ref{Fig1}, left panel). Numbers of events in the direction of
the source $N_{\rm on}$ together with the numbers of events in a region of
the sky away from the source $N_{\rm off}$ characterizing the background
are listed for each energy bin in Table.~3 of
Ref.~\cite{Aharonian:2004gb}. 
The method of gamma--hadron separation used in the HEGRA analysis does
not include cuts on $X_{\rm max}$; conservatively, we 
take the depth of the atmosphere at the HEGRA location
(approximately $1000\,\mbox{g}\,\mbox{cm}^{-2}$ for showers from the zenith
angle $\sim \, 45^\circ$) as the limiting shower depth
$X_{\rm max}^{\rm lim}$.

Data in the highest energy bin have the strongest power in
constraining LV. We apply the likelihood ratio method 
\cite{Li:1983fv,Rolke:2004mj} to these
data to test the one parameter family of LV hypotheses parameterized
by $M_{LV,\g}$. The observed values $(N_{\rm on}=36,N_{\rm off}=104)$ are
assumed to be random realizations of Poisson distributions with the
average values 
\be
\label{NsNb}
\langle N_{\rm on}\rangle
=\langle N_{\rm s}\rangle+\langle N_{\rm b}\rangle~,~~~~~~
\langle N_{\rm off}\rangle=\alpha^{-1}\langle N_{\rm b}\rangle\;,
\ee 
where $\langle N_{\rm s}\rangle$, $\langle N_{\rm b}\rangle$ are the
expectation values of the signal and background respectively, and
$\alpha=0.2$ is the ratio of the on- and
off-exposures~\cite{Aharonian:2004gb}. The expectation value of the
signal in the presence of LV is given by,
\be
\label{NsLV}
\langle N_{\rm s}\rangle^{LV}=P_{\rm reg}(E_{\rm max})\,
\langle N_{\rm s}\rangle^{LI}\;,
\ee
where $\langle N_{\rm s}\rangle^{LI}$ is the expectation value of the
signal in the standard LI theory; it is obtained by extrapolating the
flux from energies below $20$\,TeV with a power-law. The expectation
value of the background $\langle N_{\rm b}\rangle$ is unknown and is
marginalized over. The likelihood is calculated
as the probability to have the observed realization 
$(N_{\rm on},N_{\rm off})$ for a given value of $M_{LV,\g}$, 
normalized to the maximal value of the
probability over all possible choices of $M_{LV,\g}$. It is known that
the logarithm of the likelihood, multiplied by $(-2)$, obeys
the $\chi^2$ distribution.

\begin{figure}[t]
\hspace{-9mm}
\includegraphics[width=0.6\linewidth]{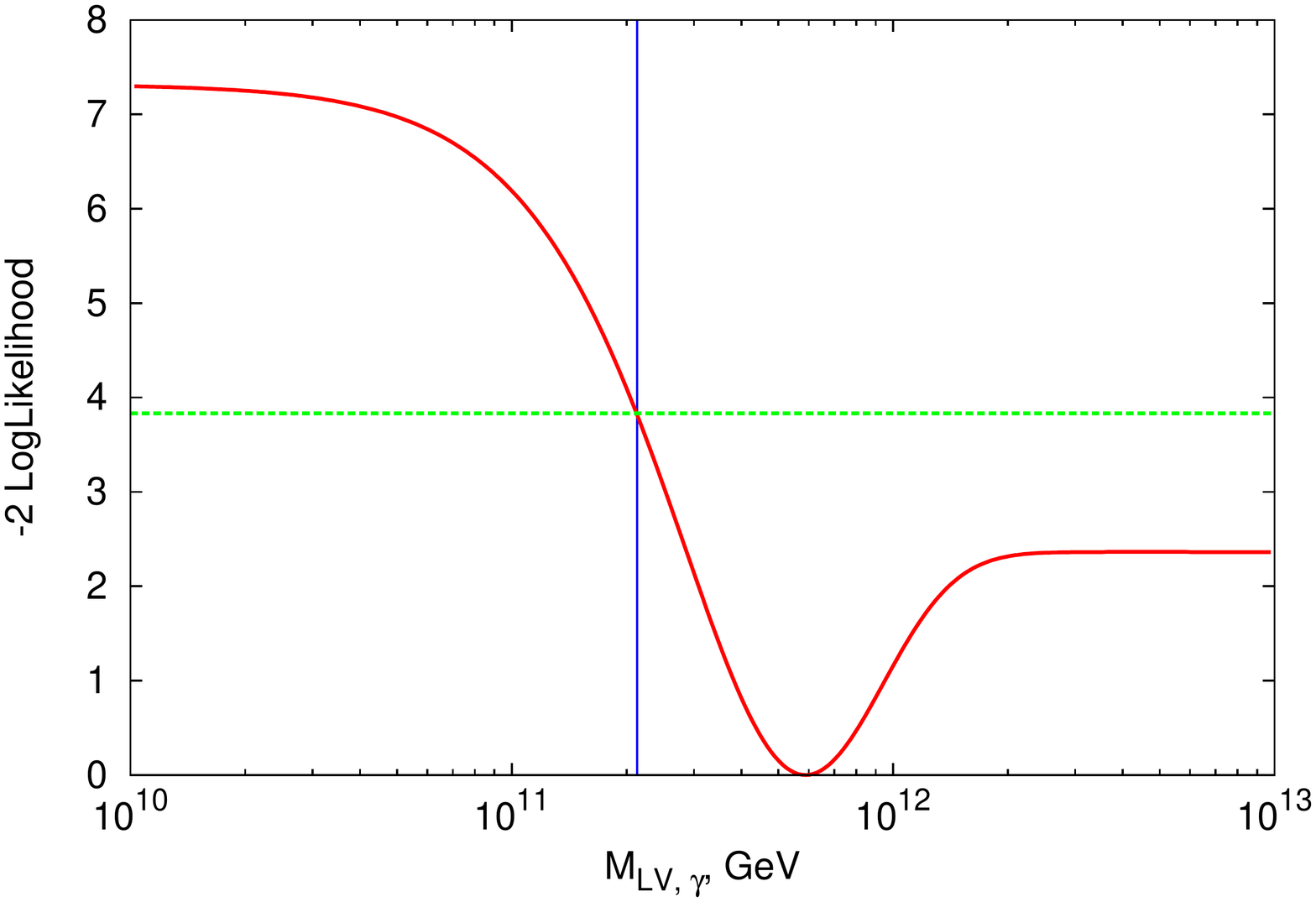}\hspace{-13mm}
\includegraphics[width=0.6\linewidth]{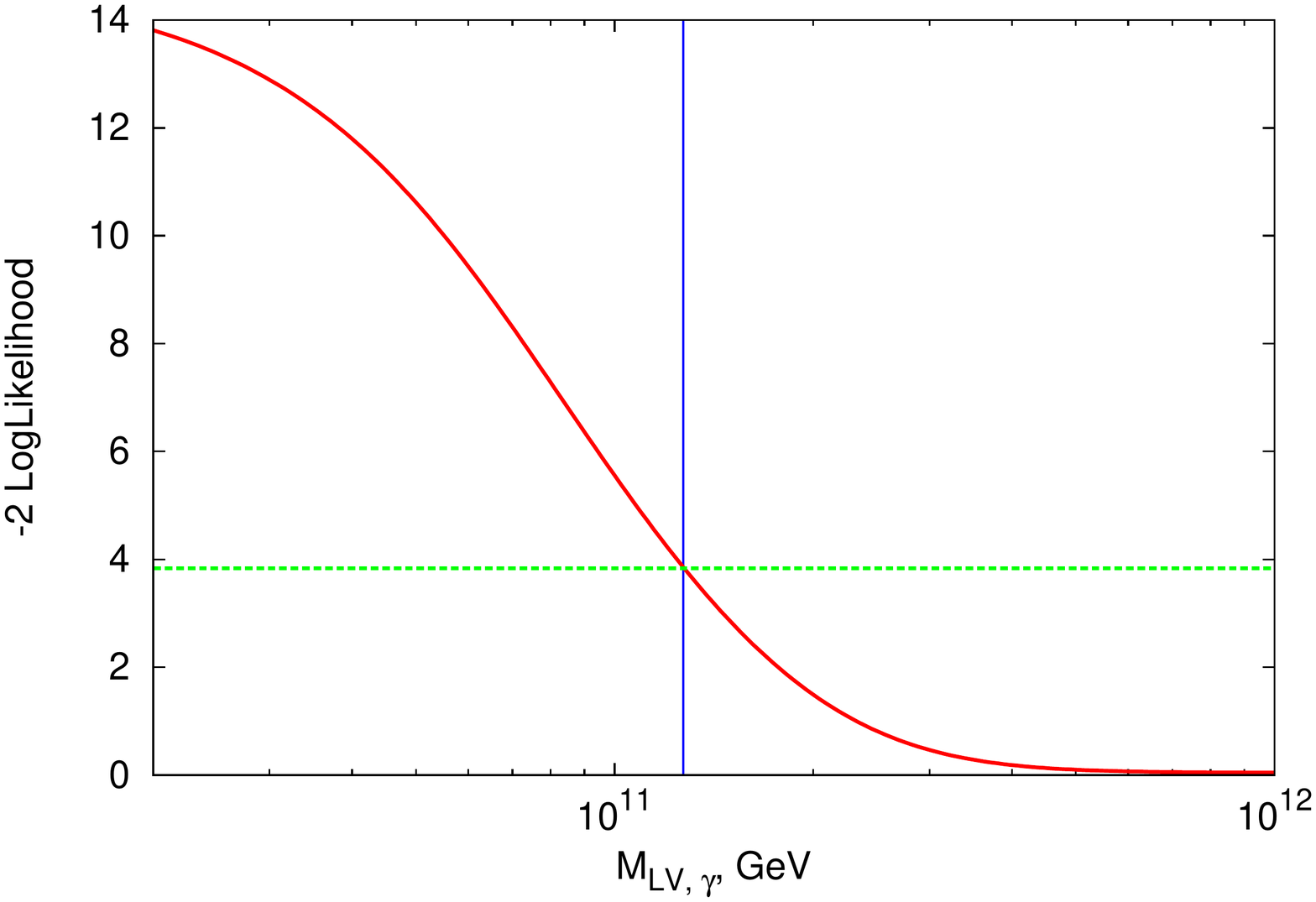}
\caption{\label{Fig2}
Dependence of the likelihood on the scale of Lorentz violation in photons
$M_{LV,\g}$ obtained using the Crab spectrum measurements by  
HEGRA (left) and H.E.S.S. (right). The values of $M_{LV,\g}$ to the
left of the vertical line are excluded at $95\%$\,CL. 
} 
\end{figure}

The resulting likelihood profile is shown in Fig.~\ref{Fig2}, left
panel. From it one reads the constraint 
\bseq
\label{hegrabound}
\begin{align}
\label{hegraboundM}
&M_{LV,\g}>2.1\times 10^{11}\,\mbox{GeV}\qquad (\epsilon_\g=-1) \qquad
&\text{at $95\%$ CL.}
\end{align}
In the effective field theory parameterization 
of \cite{Kostelecky:2009zp} this translates into a one-sided bound on the
coefficient $c^{(6)}_{(I)00}$,
\begin{align}
\label{hegraboundc}
&c^{(6)}_{(I)00}<4\times 10^{-23}\,\mbox{GeV}^{-2}
&\text{at $95\%$ CL.}
\end{align}
\eseq
The data exhibit a slight preference for 
$M_{LV,\g} \approx 6\times 10^{11}$\,GeV, but it is not statistically
significant. It is due to the fact that the observed flux
in the last bin lies below the best power-law fit. 

Let us comment on the sensitivity of the bound (\ref{hegrabound}) to
the assumptions about the intrinsic spectrum of the Crab nebula. If
instead of a pure power-law, we use a model with a cutoff in the
intrinsic spectrum, the bound on LV scale will become
stronger. Indeed, the slight steepening of the observed spectrum in
the last bins will be accounted for by the intrinsic cutoff, leaving
no room for an additional suppression due to LV. On the other hand, if
the model for the intrinsic spectrum includes hardening at high
energies, the bound on $M_{LV,\g}$ will get weaker. However, this
scenario is disfavored according
to the present theoretical understanding of the VHE photon emission in
the Crab nebula~\cite{Meyer:2010tta}. 
Thus, the
bound (\ref{hegrabound}) can be considered as conservative.

\paragraph{B. H.E.S.S. data.}
The second data sample that we consider in this paper is the
measurement of the Crab spectrum resulting from 4.4 hours of data
taking by the H.E.S.S. observatory during the flare in 
March 2013~\cite{Abramowski:2013qea}. The data extend till 
$E_{\rm max}\sim 40$\,TeV, see Fig.~\ref{Fig1}, right
panel\footnote{
The H.E.S.S. data on the Crab spectrum in the quiet
  state~\cite{Aharonian:2006pe}, 
despite being collected over a longer observation time (22.9
  hours), terminate at a lower energy $E_{\rm max}\sim 30$\,TeV and
  thus are less suitable for constraining LV.}. The
number of on- and off-events in the last bin are 
$(N_{\rm on}=4,N_{\rm off}=1)$. 
The gamma-hadron separation technique implemented in H.E.S.S. uses a
multivariate analysis method which includes, in particular, cuts on 
$X_{\rm max}$ \cite{gammashower1,gammashower2}. Conservatively, we take 
$X_{\rm max}^{\rm lim}=600\, \mbox{g}\, \mbox{cm}^{-2}$: deeper showers
certainly would not be recognized as photon events.

We use the same approach as in the case of HEGRA dataset to determine
the likelihood of the LV hypothesis. The ratio between the on-
and off-exposures for the last bin is taken as
$\alpha=0.095$~\cite{private}. The resulting likelihood curve is shown
in the right panel of Fig.~\ref{Fig2}. It implies the bound,
\bseq
\label{hessbound}
\begin{align}
\label{hessboundM}
&M_{LV,\g} > 1.3\times 10^{11}\, \mbox{GeV}\qquad (\epsilon_\g=-1) \qquad&
\text{at $95\%$ CL}\;,
\end{align}
or 
\begin{align}
\label{hessboundc}
&c^{(6)}_{(I)00} < 10^{-22}\, \mbox{GeV}^{-2}&
\text{at $95\%$ CL}\;
\end{align}
\eseq
in the notations of \cite{Kostelecky:2009zp}.
This constraint is weaker than (\ref{hegrabound}), which is a
consequence of the lower statistics. A full
statistical analysis of the H.E.S.S. data on the Crab flare including
determination of $X_{\rm max}$ for individual events has potential to
improve the bound (\ref{hessbound}). Such analysis would require an access
to the raw experimental data and is beyond the scope of the present work. \\

The constraints (\ref{hegrabound}), (\ref{hessbound}) obtained
in this subsection are of the same order, but somewhat 
weaker than the bounds from the
absorption on EBL (\ref{pp0bound}), (\ref{ppbound}). 
Still, they are important to validate the latter
bounds which rely on an implicit assumption that LV does not modify
shower formation for photons with energies below $\sim 20$\,TeV. Our
study implies that this assumption is indeed correct.

\subsection{Estimates for future experiments}
It is interesting to analyze how the bounds on LV can be improved by
future observations. Cherenkov Telescope Array (CTA)~\cite{cta} will
be able to measure the photon flux from the Crab nebula at energy
$\sim 100$\,TeV upon 50 hours of data taking for any realistic model
of the Crab emission spectrum. This forecast assumes the quality
requirements of no less than 10 signal events in each energy bin and
the statistical significance of non-zero flux detection at least
$5\sigma$~\cite{ctamonte}. To estimate the CTA sensitivity to LV we
take several sample values of $(N_{\rm on},N_{\rm off})$ that satisfy
these requirements (for $\alpha=0.2$), see Table~\ref{Tab1}. 
\begin{table}[h]
\begin{center}
\begin{tabular}{|c|c|c|c|} 
\hline
$N_{on}$ & $N_{off}$ &$\langle N_{\rm s}\rangle^{LI}$ & 95\% CL bound on
$M_{LV,\g}$ (GeV)\\ 
\hline
11 & 3  &10.4& $1.72\times 10^{12}$  \\
20 & 18 &16.4& $1.90\times 10^{12}$ \\
30 & 42 &21.6& $1.95\times 10^{12}$ \\
\hline
\end{tabular}
\end{center}
\caption{\label{Tab1} CTA exclusion potential for several realizations
of the number of events corresponding to $5\sigma$ detection of the
photon flux in the energy bin centered at 100\,TeV.}
\end{table}
Next, we assume that the expectation value of the signal predicted by
the LI model $\langle N_{\rm s}\rangle^{LI}$ coincides with the
best-fit value following from the measurements. Finally, we calculate
the likelihood dependence on $M_{LV,\g}$ following the approach
described in the previous subsection; we assume that the central energy in the bin is
 $100$\,TeV  and
use 
$X_{\rm max}^{\rm lim}=600\,\mbox{g}\,\mbox{cm}^{-2}$. The resulting
$95\%$\,CL bounds are listed in the Table~\ref{Tab1}. Conservatively,
we conclude that a $5\sigma$ detection of $100$\,TeV photon flux by
CTA will allow to constrain 
\be
\label{CTABH}
M_{LV,\g} > 1.7\times 10^{12}\, \mbox{GeV}\;\qquad (\epsilon_\g=-1) \;.
\ee
This is almost an order of magnitude stronger than the limit
(\ref{hegraboundM}) from
HEGRA data and exceeds the best current limit (\ref{ppbound}) 
by a factor of $2.5$. Similar results can be obtained in the case of a 
$100$\,TeV photon flux detection by the HAWC experiment~\cite{HAWC}.

The above analysis provides a simple criterion to estimate the
exclusion power of a given experiment that can be used also at higher
photon energies. Under the condition of a $5\sigma$ detection of the
photon flux, the values $M_{LV,\g}$ can be excluded at $95\%$\,CL if
they suppress the registration probability of the photon (\ref{Preg}) 
by at least
a factor of two,
\be
\label{Pcriter}
P_{\rm reg}(E_\g)\leq \frac{1}{2}\;.
\ee  
Extensive air shower arrays, 
such as LHAASO~\cite{DiSciascio:2016rgi}, 
TAIGA (HISCORE)~\cite{Budnev:2016btu}  
and Carpet-2~\cite{Dzhappuev:2015hxl} 
are designed to register photons with energies up to 
(a few)$\times 10^2$\,TeV. If the Crab spectrum does not have a sharp
cutoff up to these energies, they will be able to detect the
corresponding flux with high significance. Assuming a $5\sigma$
detection of photons with energies $\sim 400$\,TeV and using
(\ref{Pcriter}) as the exclusion criterion, one obtains the lower
bound,
\be
\label{TAIGABH}
M_{LV,\g} \gtrsim 3\times 10^{13}\, \mbox{GeV}\qquad (\epsilon_\g=-1) \;.
\ee

Clearly, the constraint on LV will get even stronger if photons with
yet higher energies are observed. At present we do not know if sources
of such photons exist in the universe. One possibility could be
photons produced by the interaction of UHE cosmic rays with CMB. These
photons would have energies $10^{19}\div 10^{20}$\,eV, but their flux
is highly uncertain depending on the chemical composition of UHE
cosmic rays and the unknown radio background. Nevertheless, with an
appropriate reconstruction of $X_{\rm max}$ for individual events, the
bounds on LV can be obtained without any assumptions about the origin
of primary photons or their flux, the only requirement being a
detection of a few photon-induced showers. In particular, a handful of
5 photon events with energies $\sim 10^{19}$\,eV will be sufficient to
set strong trans-Planckian constraint 
$M_{LV,\g}\gtrsim 4\times 10^{23}$\,GeV~\cite{Rubtsov:2013wwa}.  

\section{Conclusions}
\label{sec:4}

We have shown that in LV QED the cross section of the Bethe--Heitler
process responsible for the first interaction of VHE photons with the
atmosphere is suppressed compared to the LI theory. This increases the
depth of the photon-induced showers which, in turn, leads to
suppression of the number of registered VHE photons. Using absence of
such suppression in the high-energy part of the Crab spectrum we
obtained 95\% CL lower bounds on the scale of LV in photons. The bound
following from the data collected by the HEGRA experiment
(\ref{hegrabound}) is stronger than the one obtained using the
H.E.S.S. data on the Crab March 2013 flare (\ref{hessbound}), which is
due to higher HEGRA statistics. A more detailed statistical analysis
involving the characteristics of observed showers (in particular, the
values of $X_{\rm max}$) would plausibly improve the H.E.S.S. bound.

The constraints obtained in this work are a few times weaker than the
bounds derived from VHE photon absorption on EBL. Still, they play the
role of validating the latter bounds which were obtained under the
assumption of the standard shower formation probability.

We have analyzed the potential of future experiments such as CTA and
extensive air shower arrays to improve the bounds on LV from shower
formation. We have found that, depending on the maximal energy of
detected photons, the constraints can be improved by one 
($E_{\rm max}\sim 100$\,TeV) or two ($E_{\rm max}\sim 400$\,TeV)
orders of magnitude. This is comparable to the bound that can be
obtained by CTA using the EBL absorption feature in the spectrum of
Mrk 501 under the most favorable assumption of the power-law emission
spectrum\footnote{In the case of a cutoff in the emission spectrum the
EBL absorption bound is weaker.}~\cite{Fairbairn:2014kda}.

It is worth emphasizing that the bounds derived from the shower
formation mostly rely on the physical processes happening in the
atmosphere and thus are very robust. The only modeling of the source
that enters into our analysis is an assumption that a power-law spectrum
sets an upper limit on the primary photon flux. Future observations
may require more detailed models of the emission spectrum. However,
given that the most plausible source of photons with energies 
(a few)$\times100$\,TeV is the well understood Crab nebula, 
this does not
appear problematic. Moreover, a proper reconstruction of $X_{\rm max}$ for
individual events allows to get rid of any assumptions about the
primary flux, making the bounds completely independent of the source
model~\cite{Rubtsov:2013wwa}. 
 
\paragraph*{Acknowledgments}
We thank Arnim Balzer, Dieter Horns and Kornelia Stycz for 
explanations concerning the H.E.S.S. results. We are grateful to
Sergey Troitsky and Ksenia Ptitsyna for helpful discussions. 
We thank Alan Kostelecky for useful comments on the first version of
the paper.
The comparison of H.E.S.S. and HEGRA observations with the prediction of
LV models was supported by the Russian Science Foundation, grant 
14-12-01340. P.S. thanks CERN Theory Department for hospitality. The work
of S.S. is supported by the Swiss National Science Foundation.

\end{document}